\documentclass[modern]{aastex62}

\usepackage{amsmath}

\newcommand{\package}[1]{\textsl{#1}}
\newcommand{\gaia}{\textsl{Gaia}}
\newcommand{\pans}{\textsl{Pan-STARRS}}

\newcommand{\msun}{\textrm{M}_\odot}
\newcommand{\kpc}{\textrm{kpc}}
\newcommand{\kms}{\ensuremath{\textrm{km}~\textrm{s}^{-1}}}
\newcommand{\bs}[1]{\boldsymbol{#1}}
\newcommand{\masyr}{\ensuremath{\textrm{mas}~\textrm{yr}^{-1}}}
\newcommand{\feh}{\ensuremath{[\textrm{Fe} / \textrm{H}]}}
\newcommand{\article}{\textsl{Letter}}
\newcommand{\given}{\,|\,}

\newcommand{\sectionname}{Section}

\renewcommand{\tablename}{Table}

\newcommand{\changes}[1]{{#1}}

\shorttitle{GD-1 in Gaia DR2}
\shortauthors{Price-Whelan \& Bonaca}

\begin{document}\sloppy\sloppypar\raggedbottom\frenchspacing % trust me

\title{Off the beaten path: \\
       Gaia reveals GD-1 stars outside of the main stream}
% Gaia DR2 reveals stream dmlocal/tex/stars...
% A first look at the GD-1 stellar stream with Gaia DR2

\author[0000-0003-0872-7098]{Adrian~M.~Price-Whelan}
\altaffiliation{These authors contributed equally to this work.}
\affiliation{Department of Astrophysical Sciences,
             Princeton University, Princeton, NJ 08544, USA}
\email{adrn@astro.princeton.edu}
\correspondingauthor{Adrian M. Price-Whelan}

\author[0000-0002-7846-9787]{Ana Bonaca}
\altaffiliation{These authors contributed equally to this work.}
\affil{Harvard--Smithsonian Center for Astrophysics, Cambridge, MA 02138, USA}

\begin{abstract}\noindent % trust me
Tidally-disrupted globular clusters are transformed into thin, dynamically-cold
streams of stars that are extremely valuable tracers of the large- and
small-scale distribution of mass in the Galaxy.
Using data from the \gaia\ second data release combined with
\pans\ photometry, we present a sample of highly-probable members of the
longest cold stream in the Milky Way, GD-1.
The resulting map of GD-1: (1) extends the apparent length of the stream by
$20^\circ$, (2) \changes{reveals plausible locations for the progenitor}, (3) detects high-contrast gaps along the stream, and (4) indicates the existence of stream members perturbed off the main stream track.
These discoveries are only possible because of the exquisite astrometry
from \gaia, which permits a clean separation of the stream from Milky Way
stars.
The additional length and a proper treatment of the progenitor will aid in dynamical modeling of GD-1 for mapping the large-scale dark matter distribution.
The complex morphology of the stream points to a turbulent history; detailed phase-space properties of the perturbed stream members could potentially constrain dark matter substructure in the Milky Way.

\end{abstract}


\section{Introduction}
\label{sec:intro}

Stellar streams are formed during the tidal disruption of stellar systems by the
gravitational field of their host galaxy.
The phase-space density and mean track of stars in streams therefore encode
information about the underlying distribution mass on galactic
scales (e.g., \citealt{Johnston:1999, Bonaca:2018}).
% Most stellar streams are found in the halos of galaxies and are therefore
% important tracers of the large-scale distribution of dark matter in galactic
% halos.
% Long, thin stellar streams are also excellent laboratories for studying
% small-scale structures in the mass distribution:
% thin streams can remain coherent for tens of orbital periods, depending in
% detail on the symmetries of the underlying galactic mass distribution (e.g.,
% \citealt{Erkal:2016a}).
Dynamically-cold stellar streams formed from disrupted globular clusters can
also retain imprints on their phase-space density from encounters with a massive
perturber for long times \citep[e.g.,][]{Yoon:2011}.
% Encounters between massive perturbers and stream stars imprint variations in the
% phase-space density that can persist and remain visible for close to their total
% survival time (e.g., \citealt{Yoon:2011}).
In this sense, streams represent one of the most promising directions for
testing the existence of small-scale dark matter sub-halos, predicted by
standard $\Lambda$ Cold Dark Matter ($\Lambda$CDM) theory
(\citealt{Erkal:2015, Sanders:2016, Bovy:2017}).

Over 30 candidate stellar streams have been discovered in the Milky Way \citep{Grillmair:2016, Newberg:2016, Malhan:2018}.
% thanks to large-area, multi-band photometric surveys (especially
% the Sloan Digital Sky Survey, SDSS; \citealt{York:2000}).
% The Milky Way streams have a wide range of properties such as length, velocity
% dispersion, and density (e.g., \citealt{Odenkirchen:2001, Grillmair:2006,
% Grillmair:2006b, Belokurov:2006, Belokurov:2007, Bonaca:2012, Shipp:2018}; see
% also the recent review \citealt{Grillmair:2016, Newberg:2016}).
% , and the majority
% have no obvious, bound progenitor systems;
% of the thin, likely globular-cluster-origin streams, only two have clear
% progenitors (NGC 5466 and Palomar 5).
% Most streams were first detected as a linear overdensity in stellar counts filtered to preferentially select faint blue stars, most likely members of a former Galactic satellite.
The GD-1 stream is the most prominent among the thin streams, discovered as an
overdensity of faint blue stars in the SDSS photometry \citep{Grillmair:2006}.
Initially detected to span $\sim 60^\circ$ on the sky, the physical length of GD-1 at a distance of $\sim
7$--$10~\textrm{kpc}$ is at least $\sim 15~\kpc$.
No remnant progenitor for the stream has been found, but its width ($\sigma
\approx 12'$), metallicity ($\feh \approx -1.4$), and estimated stellar mass ($M
\approx 2 \times 10^4~\msun$) imply that GD-1 is a disrupted globular cluster
\citep{Koposov:2010}.

Its length and location in the Galactic halo (pericentric distance
$r_\textrm{peri} \sim 13~\kpc$) make GD-1 an ideal object for constraining dark
matter properties in the inner Milky Way.
Both from fitting orbits to binned phase-space measurements along the GD-1
stream (\citealt{Koposov:2010}) and from modeling the stream track in
phase-space coordinates \citep{Bowden:2015, Bovy:2016}, it appears that the dark
matter distribution within $\approx20\,\rm kpc$ is close to spherical, consistent
with findings from simulated Milky Way-like galaxies \citep{Zhu:2016}.
% This is consistent with findings from simulated Milky Way-like galaxies:
% simulated dark-matter-only halos tend to be triaxial \citep{Allgood:2006}, but
% baryonic processes tend to make their inner regions more spherical
% \citep{Zhu:2016}.

Its implied old dynamical age, combined with its orbital properties, makes GD-1
a prime stream to also study the small scale structure of dark matter.
While density variations in streams can be induced from interactions with giant
molecular clouds \citep{Amorisco:2016} or resonant encounters with the bar
\citep{Pearson:2017}, these baryonic effects are unlikely to affect GD-1 because
of its large pericentric distance and retrograde orbit with respect to the
Galactic bar.
Therefore, observed density variations in the GD-1 stream could instead be an
indication of past interactions with dark matter subhalos
\citep[e.g.,][]{Ngan:2014}.
% By estimating a stream age of $\approx 4~\textrm{Gyr}$, assuming a progenitor
% mass of $10^5~\msun$, and by assuming properties of the subhalo population from
% dark matter-only simulations \citep{Springel:2008, Diemand:2008},
\citet{Erkal:2016} predict that a 4\,Gyr old GD-1-like stream in a Milky Way-like galaxy
could have up to one significant, wide
($\approx 5$--$7^\circ$) gap caused by an interaction with a
$10^6$--$10^7~\msun$ subhalo.
A definitive ruling on the presence of dark matter subhalos in this mass regime
would directly inform the nature of dark matter \citep[e.g.,][]{Bullock:2017}.
% dark matter-only simulations \citep[scaled to the Milky Way halo mass and
% reduced in number by a factor of 3 to account for disruption by the Galactic
% disk]{Springel:2008, Diemand:2008}, \citet{Erkal:2016} predict that GD-1 could

Photometric studies of GD-1 revealed the presence of density variations and gaps
along the stream, as well as wiggles in the main stream track
\citep[][]{Carlberg:2013, DeBoer:2018}.
However, purely photometric studies of streams suffer from contamination of the
Milky Way foreground, which, on small scales, could partially account for
density inhomogeneities reported along cold streams \citep[e.g.,][]{Ibata:2016}.
In this \article, we improve upon the selection of likely GD-1 members by using
astrometric data from the \gaia\ mission, combined with precise photometry from
the \pans\ survey (\S\,\ref{sec:data}).
These data enable us to measure density variations along GD-1 to an
unprecedented precision (\S\,\ref{sec:res_global}), and detect clear signatures
of stream stars beyond the main track (\S\,\ref{sec:res_gap}).
Implications of this revolutionary remapping of GD-1 are discussed in
\S\,\ref{sec:discussion}.
% \todo{hints of density variations} from initial work Koposov:2010, later confirmed by density modeling, but all low significance
% (\citealt{Carlberg:2013}).
% More recently, deep photometry of GD-1 from CFHT/Megacam has revealed
% interesting small-scale ``wiggles'' and significant density variations along a $\approx 45^\circ$ span of the stream (\citealt{DeBoer:2018}):
% [summarize their conclusions].
% \todo{these analyses relied on binned data, still contaminated by MW stars}
%
% In this \article, we present clearest view of the GD-1 stream to date using
% astrometric data from the \gaia\ mission data release 2 (\DR) combined with
% precise photometry from the \pans\ survey.
% We find XX under-densities gaps, some consistent with ...
% We revisit...do not try to model or fit for Galactic potential.
% Focus on stream properties, data.

\section{Data}
\label{sec:data}

We use astrometric data from the \gaia\ mission (\citealt{Prusti:2016}), data
release 2 (\citealt{Gaia-Collaboration:2018, Lindegren:2018}), and photometry
from the \pans\ survey, data release 1 (\citealt{Chambers:2016}) to select
high-confidence members of the GD-1 stream.

\begin{figure}
\begin{center}
\includegraphics[width=\textwidth]{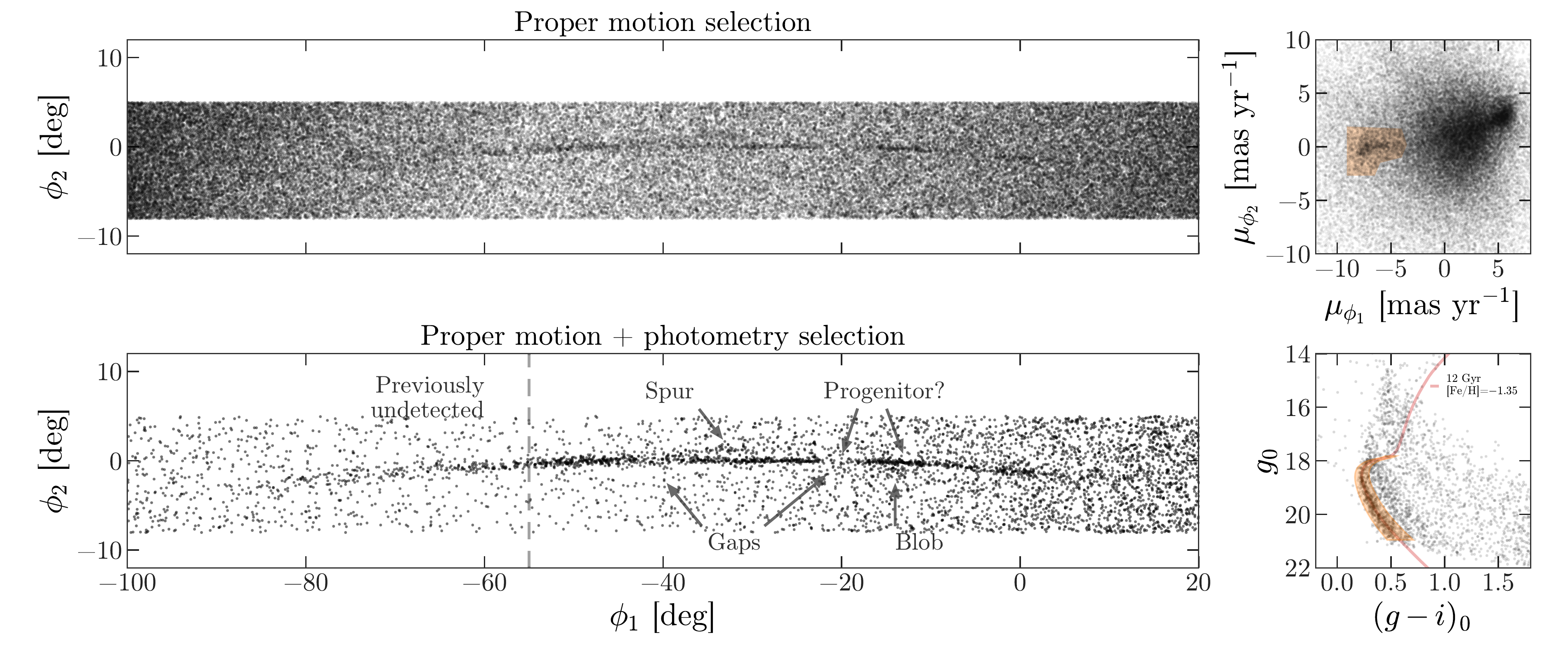}
\end{center}
\caption{
On-sky positions of likely GD-1 members in the GD-1 coordinate system.
GD-1 is apparent as an overdensity in negative proper motions (top right panel,
orange box), so selecting on proper motion already reveals the stream in
positions of individual stars (top left).
The stream also stands out in the color-magnitude diagram (bottom right) as
older and more metal poor than the background.
Selecting the main sequence of GD-1 (orange, shaded region in bottom right)
along with proper motion cuts unveils the stream in unprecedented detail (bottom
left).
% This map extends the stream by $20^\circ$, identifies a plausible location of the progenitor at $\phi_1\sim13^\circ$, confirms the existence of a significant gap at $\phi_1\sim-20^\circ$, shows additional under-densities at $\phi_1\sim-40^\circ,-5^\circ$ and for the first time finds stream overdensities off from the main track of a cold stream at $(\phi_1,\phi_2)\sim(-35^\circ,-1^\circ)$ and $(-15^\circ,-2^\circ)$.
}
\label{fig:selection}
\end{figure}

We retrieve data along the previously-identified track of the GD-1 stream
from the \gaia\ science archive\footnote{\url{https://gea.esac.esa.int/}} by
selecting sources with small parallax, $\varpi < 1~\textrm{mas}$ in fields along the GD-1 stream.
% , and select
% batches of stars that have small latitude in the heliocentric GD-1 coordinate
% system, $(\phi_1, \phi_2)$, defined in \cite{Koposov:2010}.
We convert the equatorial sky coordinates and proper motions %of the parallax-selected stars
to the GD-1 coordinate system \citep[$\phi_1, \phi_2$,][]{Koposov:2010}, and correct the proper motions for solar reflex
motion by assuming that stars at a given stream longitude, $\phi_1$, have a
distance given by $d(\phi_1) = (0.05 \, \phi_1 + 10)~\textrm{kpc}$;
we assume a solar velocity $\bs{v}_\odot = (11.1, 232.24, 7.25)~\kms$
(\citealt{Schonrich:2010, Bovy:2015}).
\changes{In detail, we project the solar velocity vector onto the tangent space at each star's sky position, convert to angular motion using the assumed distance, and subtract the solar component from each star's proper motion (this is done using the \package{astropy} \citep{astropy} and \package{gala} \citep{gala} packages).}
\figurename~\ref{fig:selection} (top right) shows stars with $|\phi_2| <
1^\circ$ in solar-reflex-corrected proper motion components in the GD-1 system,
$\mu_{\phi_1, \odot}$ and $\mu_{\phi_2, \odot}$.
GD-1 members are visible as the overdensity around $(\mu_{\phi_1, \odot},
\mu_{\phi_2, \odot}) \approx (-8, 0)~\masyr$.

As a first selection of GD-1 member stars, we choose stars \changes{that lie in a polygon in proper motions (orange shaded polygon in top right panel of \figurename~\ref{fig:selection}; approximately bounded by the rectangle
$-9 < \mu_{\phi_1, \odot} < -4.5~\masyr$ and $-1.7 < \mu_{\phi_2, \odot} <
1~\masyr$).}
\figurename~\ref{fig:selection} (top left) shows all stars that pass the
selection, plotted in the GD-1 coordinate system.
The stream is already identifiable as the overdensity of stars around $\phi_2 =
0$ between longitudes $-60^\circ \lesssim \phi_1 \lesssim 0^\circ$.

To improve the contrast of the stream over the background, we cross-match the
sample to the \pans\ photometric catalog. % and use the photometry to clean the sample.
\figurename~\ref{fig:selection} (bottom right) shows the proper-motion-selected
candidate stream stars within $\left|\phi_2\right| < 1^\circ$ in a \pans\
color-magnitude diagram (CMD) de-reddened following \citet{Schlafly:2011}.
The GD-1 stellar population stands out from the Milky Way foreground, and is well matched by a 12\,Gyr, $\feh = -1.35$ MIST isochrone at a distance of $7.8\,\kpc$ \citep[red line in \figurename~\ref{fig:selection},][]{Dotter:2016, Choi:2016, Paxton:2011}.
% The over-plotted isochrone (red line) is from the MESA Isochrones \& Stellar
% Tracks (MIST; \citealt{Dotter:2016, Choi:2016, Paxton:2011}) and represents a
% $12~\textrm{Gyr}$ old population with $\feh = -1.35$ at a distance of
% $7.8~\kpc$.
We use this isochrone to generate a polygonal selection in de-reddened $g-i$
color and apparent $g$-band magnitude (shaded region in the
bottom right panel of \figurename~\ref{fig:selection}).

\figurename~\ref{fig:selection} (bottom left) shows the final sample of
candidate stream members after selecting on both proper motion and photometry.
The GD-1 stream is clearly visible as an overdensity of \emph{individual}
stars (the positions are not binned); this is the most pure view of the stream
to date.
This increased contrast shows that the stream extends at least another
$20^\circ$ to negative longitudes.
Furthermore, the stream reaches its highest surface density where it is the
narrowest ($\phi_1\sim-13^\circ$), which may be the location of its elusive and
fully-disrupted progenitor.
\changes{At this location, between $\phi_1 \in (-18, -10)^\circ$, we find that the mean equatorial sky position of the stream stars is $(\alpha, \delta) = (177.01, 53.99)^\circ$ and the mean proper motion is $(\mu_\alpha \, \cos\delta, \mu_\delta) = (-7.78, -7.85) \pm 0.03~\masyr$, taking into account the covariance matrix for proper motions provided with the \gaia\ data.}
\changes{The full data for the GD-1 region along with selection masks are available through Zenodo (\dataset[10.5281/zenodo.1295543]{https://doi.org/10.5281/zenodo.1295543})\footnote{See also \url{https://github.com/adrn/gd1-dr2}.}}

Several of the under-densities and gaps hinted at from photometric selection
(\citealt{Carlberg:2013, DeBoer:2018}) appear as striking features in this
significantly cleaner map of the GD-1 stream.
% Finally, at least two new features are visible near the stream: (1) ``the
% spur,'' stars above the main stream track (in $\phi_2$) between $-40^\circ
% \lesssim \phi_1 \lesssim -30^\circ$, and (2) ``the blob,'' stars below the main
% stream track between $-20^\circ \lesssim \phi_1 \lesssim -10^\circ$.
% We discuss each of these in more detail in \sectionname
% s~\ref{sec:results}--\ref{sec:discussion} below.
Finally, at least two new features are visible near the stream: (1) ``the
spur,'' stars above the main stream track (in $\phi_2$), and (2) ``the blob,''
stars below the main stream track, both highlighted in
\figurename~\ref{fig:selection}.
% We discuss each of these in more detail in \S\,
% \ref{sec:results}--\ref{sec:discussion} below.

\section{Results}
\label{sec:results}

% We extract high-confidence stream stars from the proper motion and CMD selected
% sample by further selecting stars close to the main stream track.
% We use these main stream track stars to study the global properties of the
% stream (\S\,\ref{sec:res_global}), and find interesting substructures off of the main stream track (\S\,\ref{sec:res_gap}).

\subsection{Properties of the main GD-1 track}
\label{sec:res_global}

\begin{figure}
\begin{center}
\includegraphics[width=\textwidth]{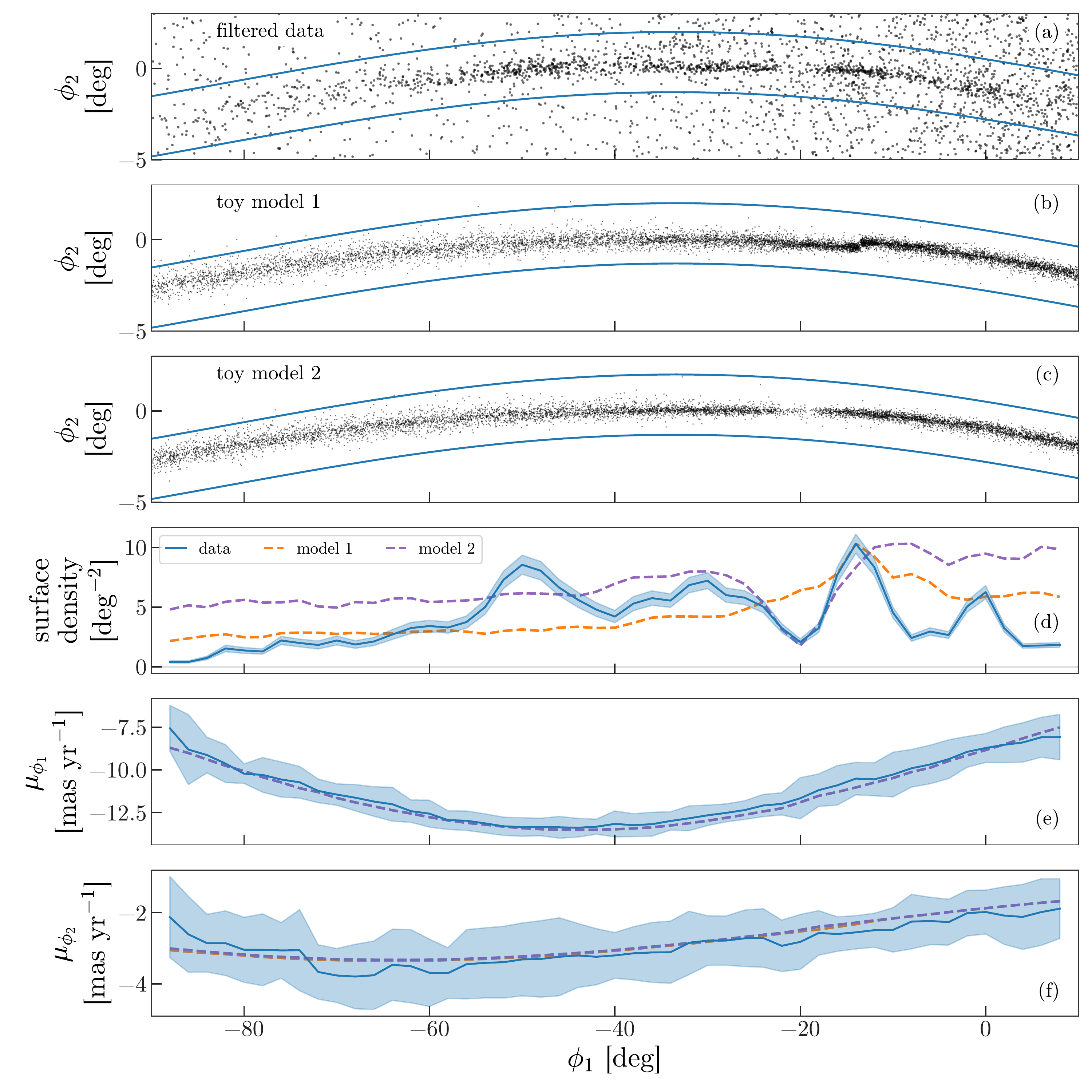}
\end{center}
\caption{%
Median stream properties computed from the proper motion and CMD filtered GD-1
stars, and the same for a toy model for the stream track.
\textit{Panel (a)}: Sky positions of candidate GD-1 stars.
Lines show a \changes{4th}-order polynomial fit to the median stream track offset by $\pm 0.75^\circ$; the enclosed region is adopted as the nominal stream region.
\changes{\textit{Panels (b) and (c)}: Sky positions of simulated star particles from mock stream models of the GD-1 stream in which the progenitor is currently disrupting (toy model 1, panel b) or the progenitor fully disrupted 500 Myr ago and results in the observed under-density at $\phi_1 \sim -20^\circ$ (toy model 2, panel c).}
\textit{Panel (d)}: Background-subtracted surface-density estimates computed in
successive, overlapping $3^\circ$ windows for both GD-1 data and the model
stream.
\textit{Panel (e)}: Median proper motion along the stream computed in the same
bins as previous panel \changes{(not solar reflex corrected)}.
\changes{The two toy models have consistent median proper motions (overlapping dashed lines).}
\textit{Panel (f)}: Same as Panel (d), but for proper motion in stream latitude.
}
\label{fig:track-and-model}
\end{figure}

To study the global properties of the stream as a function of stream longitude,
we extract stream stars around a fourth-order polynomial model for the stream
latitude, $\phi_2$, as a function of stream longitude, $\phi_1$.
We find the stream midpoints by computing the median $\phi_2$ in $4^\circ$
windows shifted by $2^\circ$ along $\phi_1$, then fit a parabola to the median
positions to find $\phi_{2, \textrm{track}}(\phi_1)$.
We define the stream region as $\left| \phi_2 - \phi_{2,
\textrm{track}}(\phi_1) \right| < 0.75^\circ$.
\figurename~\ref{fig:track-and-model}, panel (a), shows the high-confidence
stream stars (same as \figurename~\ref{fig:selection}), and the two curved lines
(blue) show the upper and lower boundary of the stream region.

We extract properties of the stream from the region defined above as a function of $\phi_1$ by computing median stream properties in (overlapping) $3^\circ$ windows.
% we set the window size to $3^\circ$ in $\phi_1$, and shift the window center by
% $1^\circ$ in longitude for each subsequent measurement.
Panels (c)--(e) in \figurename~\ref{fig:track-and-model} show the
background-subtracted stream surface density and median proper motion components
extracted in this way (dark lines), with the associated empirical scatters
(shaded regions).
We estimate the local background surface density as an average of areas north
and south of the bounded stream region.
The median proper motion profiles along the stream (panels (d)--(e)) are
consistent with previous mean proper motion measurements
(\citealt{Koposov:2010}), but here the proper motion gradient is resolved for
individual stars (see, e.g., elongated overdensity in upper right panel of
\figurename~\ref{fig:selection}).

Clear density variations, also apparent in the 2D positions of the sources, manifest as sharp features in the surface density estimates along the stream.
We have verified that these do not correspond to features in dust extinction nor are they caused by completeness variations in the \gaia\ scanning pattern:
the visible portions of the GD-1 stream are located at high Galactic latitudes ($b > 20^\circ$), and 99\% of our sample has at least 8 \gaia\ visibility periods \changes{and satisfy a selection on the unit weight error (see Equation~C.1 in \citealt{Lindegren:2018})}, promising a robust astrometric solution.
\changes{
In detail, we have compared the fraction of good solutions in the region near the spur, in the region between the stream and the spur, and in the stream at the same range of longitudes.
We consider the longitude range $-35^\circ < \phi_1 < -30^\circ$, and 3 ranges in $\phi_2$: $(-0.4, 0.4)^\circ$, $(0.4, 0.9)^\circ$, $(0.9, 1.7)^\circ$ for the stream, region between stream and spur, and spur, respectively.
In these 3 regions, we find 84\%, 88\%, 89\% of the stars with 5-parameter solutions pass the unit weight error criteria.
Small-scale density variations, however, especially apparent between $-80^\circ < \phi_1 < -50^\circ$, are a consequence of the \gaia\ scanning pattern.
}

% First, the visible portions of the GD-1 stream are located at high Galactic
% latitudes ($b > 20^\circ$).
%, and we therefore do not expect significant dust
% extinction or variations in extinction within the stream region.
% The median $V$-band extinction over the stream track is
% $\textrm{med}\left(A_V\right) = 0.04~\textrm{mag}$ with a dispersion
% $\sigma_{A_V} \approx 0.03~\textrm{mag}$, as computed from the SFD extinction
% map (\citealt{Schlegel:1998}).
% As the stream approaches the Galactic disk ($\phi_1 < -60^\circ$),
% dust extinction becomes appreciable, and the stream stars are harder to
% select apart from the background.
% Second, 99\% of the stars in our sample have at least 8 \gaia\ visibility
% periods (median and root-variance of 14 and 3, respectively), promising a robust
% estimate of the proper motions.
% The variance in the number density of faint stars originating from the \gaia\
% scanning pattern is only $\approx 1~\textrm{deg}^{-2}$, as estimated using stars
% with proper motions $-9 < \mu_{\phi_1, \odot} < -4.5~\masyr$ and $1 <
% \mu_{\phi_2, \odot} < 4~\masyr$, far smaller than the density variations in the
% stream.
% The relative ratios between different over- and under-densities might be affected by the \gaia\ completeness at the faint end, but they
The observed density variations are therefore likely real morphological changes
along the GD-1 stream.
% as have been suggested before photometric selection alone
% (\citealt{Carlberg:2013, DeBoer:2018}).
The deep under-densities we see from the kinematically-cleaned sample correspond
to previously reported gaps in the stream (\citealt{Carlberg:2013, DeBoer:2018}).
There is also a well-defined, sharp overdensity in the surface density profile
close to $\phi_1 \approx -13^\circ$ with roughly symmetric under-densities on
either side:
this is suggestive of a progenitor system in the final stages of dissolution
(\citealt{Balbinot:2018}).
This region is also the thinnest part of the stream, as would be expected for
recently stripped material; we defer a robust analysis of variations in the
stream width to future work.
\changes{With present data, however, we cannot rule out a scenario in which the progenitor system disrupted longer ago and resulted in a gap at $\phi_1\approx-20^\circ$ (see Figure~\ref{fig:track-and-model} and, e.g., \citealt{DeBoer:2018}).}
% that has not yet phase-mixed in multiple dimensions.

We compare the measured stream properties to a simple model for the phase-space
density of the stream, generated by simulating the orbital evolution of globular
cluster stars once they are tidally stripped from the progenitor.
We use the ``particle-spray'' stream generation method (\citealt{Fardal:2015})
and assume a uniform mass-loss history to generate the model stream \citep[e.g.,][]{Kupper:2012}.
%, which
% matches results from more detailed ($N$-body) numerical experiments of globular
% cluster disruption \citep[e.g.,][]{Kupper:2012}.
% During simulation, star particles are released from the Lagrange points of the
% cluster with a spread in position and velocity set by the mass of the
% satellite and the location within the host galaxy, with some tunable scale
% parameters (using the parametrization and adopted scale parameters of
% \cite{Fardal:2015}).  This method can accurately reproduce the mean track of
% the resulting stream, but the density of stars along the stream depends in
% detail on the mass-loss history and internal kinematics of the progenitor
% system.  Here we assume a constant mass-loss rate, and release stream
% particles uniformly in time from each Lagrange point; these choices match
% results from more detailed numerical experiments of globular cluster
% disruption \citep[e.g.,][]{Kupper:2012}.

\begin{figure}
\begin{center}
\includegraphics[width=\textwidth]{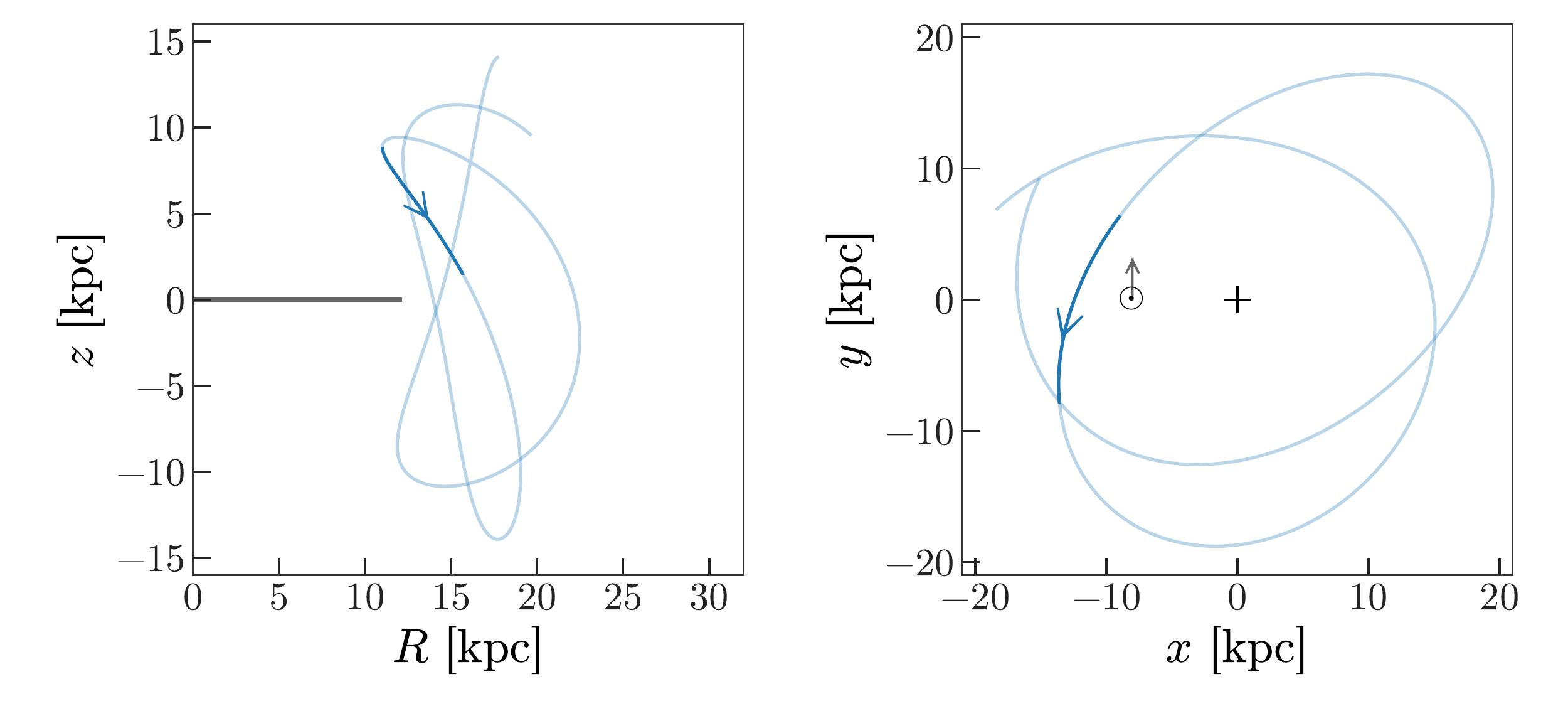}
\end{center}
\caption{%
Best-fit orbit in a simple model for the gravitational potential of the Milky
Way using \gaia\ proper motions and the filtered stream track.
Left panel shows the orbit in Galactocentric cylindrical radius, $R$, and
position above the Galactic midplane, $z$;
gray line at $z=0$ shows the approximate radius of the disk.
Right panel shows the orbit projected onto the midplane in Galactocentric
cartesian coordinates, ($x$, $y$).
Darker orbit section highlights the stream longitudes over which we clearly see
GD-1 stream stars.
Note that GD-1 is retrograde with respect to the disk (location and direction of
motion of the sun are indicated in the right panel, $\odot$).
}
\label{fig:orbitfit}
\end{figure}

To compute initial conditions for the stream model, we fit an orbit to the
observed properties of the stream in a fixed Milky Way model similar to \citet{Bovy:2015}.
% consisting of a
% disk (\citealt{Miyamoto:1975}), a bulge (\citealt{Hernquist:1990}), and a
% spherical dark matter halo (\citealt{Navarro:1996}).
% We set the disk scale-length parameters to match the disk model found in
% \citet{Bovy:2015}; other parameters are set to produce an approximately flat
% disk rotation curve with a circular velocity at the solar circle,
% $v_{\textrm{circ}, \odot} \approx 230~\kms$.
%use L-BFGS-B optimization as implemented in the \package{scipy}
% \texttt{Python} package to
We maximize the likelihood of the orbit given sky track and proper motions from
this work, with mean distance and radial velocities from
\cite{Koposov:2010}.
% We do not vary the parameters of the Milky Way model in the fit because we are
% only interested in producing a comparison stream model. % of the stream properties.

\figurename~\ref{fig:orbitfit} shows \changes{a $1\,\textrm{Gyr}$ segment of} the best-fitting orbit from the procedure outlined above.
The GD-1 stream crosses the midplane at large radius, far from the bulk of the
stellar density:
the grey line in the left panel shows the radius that contains $\approx 90\%$ of
the mass, assuming a \citet{Bovy:2012} model of the stellar disk.
% an exponential radial density profile with a scale radius
% $\rm R_d = 3~\kpc$ (an intermediate value of recent measurements, e.g.,
% \citealt{Bovy:2012}).

\changes{We present two progenitor configurations which reproduce certain aspects of the observed GD-1 density variations: the progenitor system is either at $\phi_1 = -13.5^\circ$, the peak of the observed surface density profile (toy model 1), or was fully-disrupted 500\,Myr in the past and the under-density at $\phi_1 = -20^\circ$ represents the fully-disrupted progenitor (toy model 2).}
% \changes{To initialize the model stream generation, we assume that the present day position of the progenitor system is either at $\phi_1 = -13.5^\circ$, the peak of the observed surface density profile (toy model 1), or was fully-disrupted 500 Myr in the past and the under-density at $\phi_1 = -20^\circ$ represents the fully-disrupted progenitor (toy model 2).}
We numerically integrate backwards from these locations for $4~\textrm{Gyr}$,
then forward-generate the model streams, releasing stream particles at each
timestep.
We assume an initial progenitor mass of $M=10^5~\msun$, and linearly decrease
the mass of the progenitor until its full disruption at the \changes{present (toy model 1), or at $t = -500~\textrm{Myr}$ (toy model 2)}.
% day, where we assume $M = 0~\msun$
% (i.e. fully disrupted).
Stream properties computed from the resulting model streams are plotted in
\figurename~\ref{fig:track-and-model} as dashed (orange and purple) lines.
The model streams qualitatively match the stream track in all phase-space
dimensions, but have much smoother surface density profiles along the extent of
the stream.
The broadening in the observed stream at negative longitudes is also seen in the model stream and is expected for older parts of the stream that have phase-mixed for longer times.
We therefore don't expect that the observed surface density variations can be fully explained by, e.g., epicyclic overdensities (\citealt{Kupper:2012}).

\subsection{Off-track features}
\label{sec:res_gap}

\begin{figure}
\begin{center}
\includegraphics[width=\textwidth]{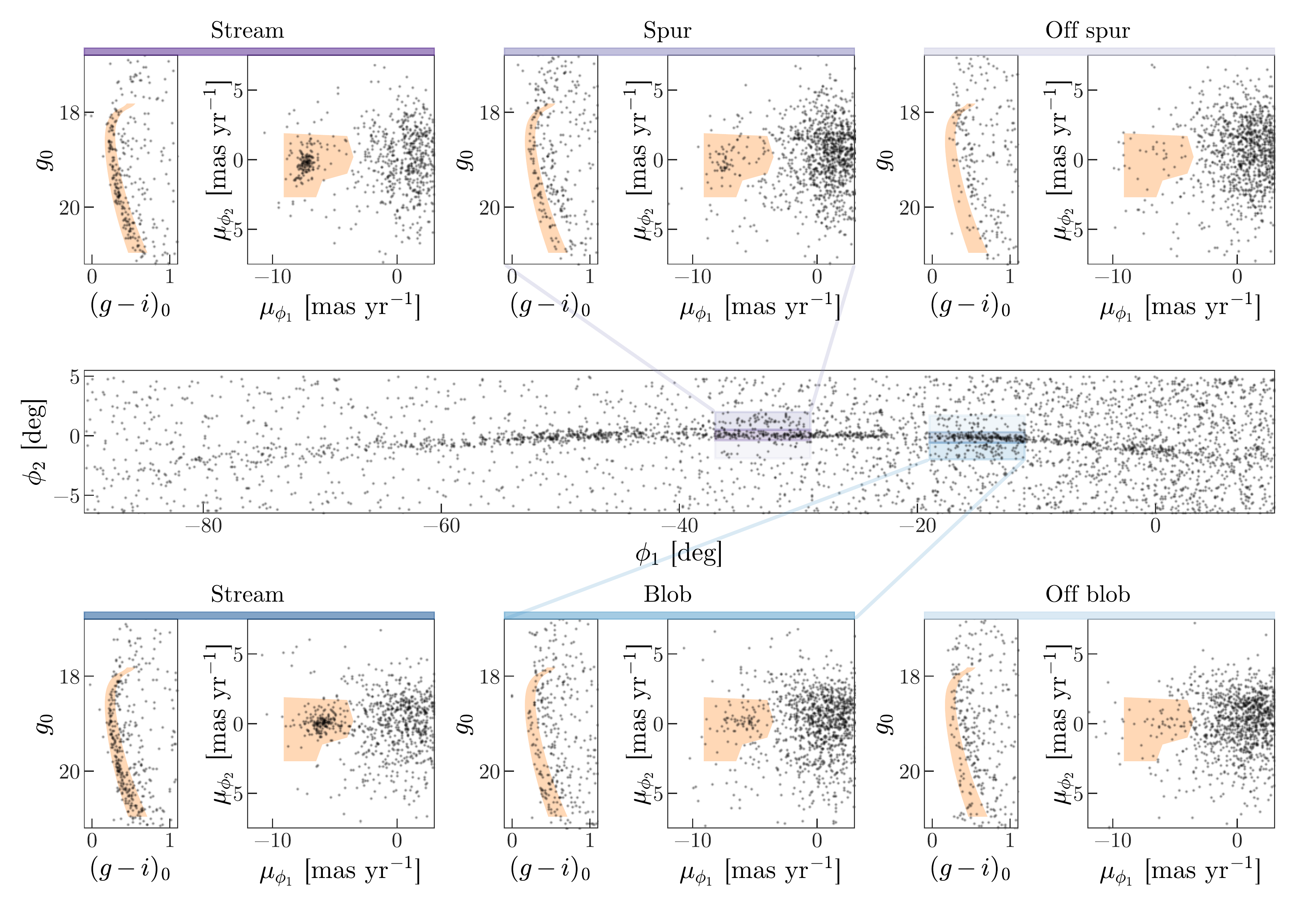}
\end{center}
\caption{
Color-magnitude and proper-motion diagrams of fields associated with a spur above GD-1 (top), and fields related to the blob below GD-1 (bottom).
Panels on the left show properties in stream fields (darkest rectangles in middle panel), middle panels present spur and blob (medium rectangles), and the right-most panels are control fields on the opposite sides from spur and blob (lightest rectangles).
% The top and bottom left panels show properties in stream fields (darkest rectangles in middle panel), middle panels present spur and blob (medium rectangles), while the right-most top and bottom panels are selected opposite from spur and blob (lightest rectangles).
Selection boxes are shown in all CMD and proper-motion panels as shaded regions.
Both the spur and the blob fields have more stars in the CMD selection box than the corresponding control fields.
Proper motions of these stars follow the distribution of stars in adjacent stream fields, thus confirming the association of these off-track features with GD-1.
}
\label{fig:features}
\end{figure}

Here we test whether the overdensities apparent beyond the main stream track
(\figurename~\ref{fig:selection}) are truly associated with the GD-1 stream by
comparing color-magnitude diagrams and proper motions of the spur and blob
fields to the stream and background fields.
% this would be a novel insight into disruption from the \gaia\ mission
% Some contamination still remains present in our selection, so it is possible that these features are due to the Milky Way population randomly over-populating our selection boxes.

The central panel of \figurename~\ref{fig:features} shows the map of GD-1, with fields of interest marked in shades of purple for spur-related fields, and shades of blue for blob-related fields.
In both cases, we analyze the stream region (darkest shade), overdensity off from the stream (medium shade) and use a field on the opposite side of the stream from the overdensity as a control (lightest shade).
The $\phi_1$ range is the same for all three fields for a given feature, and the control and feature fields have the same area (the stream field is smaller to avoid cross-contamination).

For each of the examined fields, we present the color-magnitude diagram of stars selected on proper motions (left subpanels) and proper motions of photometrically selected stars (right subpanels), with spur fields on the top and blob fields on the bottom of \figurename~\ref{fig:features}.
% In both rows, arranged from left to right are the stream, the feature, and the control field.
Stars selected from the stream fields show a clear overdensity in the CMD selection box and are clustered in proper motions.
The two stream fields have a clear relative offset in proper motions, as expected from the velocity gradients shown in \figurename~\ref{fig:track-and-model}.
Furthermore, the main-sequence turn-off of the $\phi_1\sim-15^\circ$ field appears to be fainter than the $\phi_1\sim-35^\circ$, consistent with the distance gradient across the stream measured by \citet{Koposov:2010}.

The control fields have some stars in both the CMD and proper-motion selection boxes, but without any significant overdensities in either.
CMDs of the spur and blob fields have more stars in the selection box than their respective control fields, and the distribution within the box is qualitatively similar to that of the corresponding stream field, suggesting a similar distance.
In proper motions, the spur and blob fields also display properties similar to the adjacent stream fields, with clear overdensities of stars in the selection box.

\begin{figure}
\begin{center}
\includegraphics[width=\textwidth]{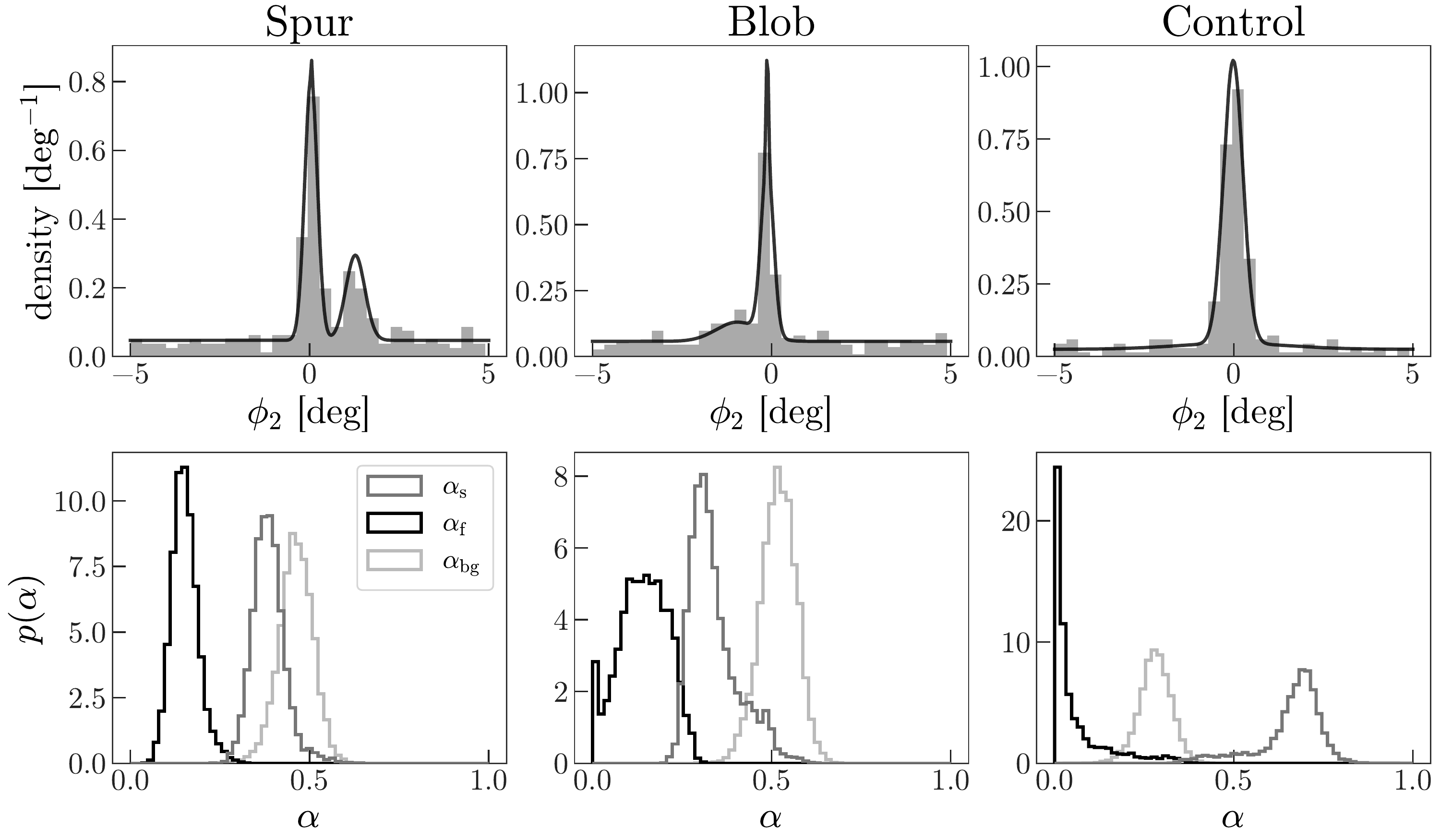}
\end{center}
\caption{
\changes{\emph{Top row}: Density of stars in $\phi_2$ (grey) and inferred density (black curve) for regions around the spur, blob, and a control field.
The model for the density includes components for the stream, background, and a feature overdensity (the spur or the blob), but with amplitudes that are allowed to go to zero (see \sectionname~\ref{sec:res_gap} for more details.)
\emph{Bottom row}: Inferred posterior probability distributions over the amplitude ($\alpha$) of each component in each field: $\alpha_{\textrm{s}}$ for the stream, $\alpha_{\textrm{bg}}$ for the background, and $\alpha_{\textrm{f}}$ for the feature.
In both feature fields, the model prefers an additional component to describe the feature, whereas in the control field the density is consistent with stream + background.}
}
\label{fig:density-model}
\end{figure}

\changes{
To quantify the significance of these features, we construct a statistical model of the linear density of the filtered sky positions as a function of $\phi_2$ around each of the features and at a control region.
We select three equal ranges of $\phi_1$ centered on the spur, $\phi_1 \in (-36, -30)^\circ$, the blob, $\phi_1 \in (-18, -12)^\circ$, and in between (as a control field), $\phi_1 \in (-51, -45)^\circ$, and model the one-dimensional density in $\phi_2$ in each of these regions. % to see if the number of stars at the location of each feature is consistent with the background level.
We use a four-component mixture model to represent the density consisting of a uniform background over the range $\phi_2 \in (-10, 5)^\circ$, a two-component Gaussian mixture to represent the main stream with mean $\mu_{\textrm{s}}$ and variances $\sigma_{\textrm{s}, 1}^2$ and $\sigma_{\textrm{s}, 2}^2$, and a single Gaussian for each feature with mean and variance $(\mu_{\textrm{f}}, \sigma_{\textrm{f}}^2)$.
The full density model given all parameters $\bs{\theta} = (\alpha_{\textrm{bg}}, \alpha_{\textrm{s}}, \mu_{\textrm{s}}, \sigma_{\textrm{s}, 1}, \sigma_{\textrm{s}, 2}, \mu_{\textrm{f}}, \sigma_{\textrm{f}})$ is
\begin{equation}
    \begin{split}
    p(\phi_2 \given \bs{\theta}) &=
            \alpha_{\textrm{bg}} \, \mathcal{U}(-10, 5) \\ & \quad +
            \alpha_{\textrm{s}, 1} \, \mathcal{N}(\phi_2 \given \mu_{\textrm{s}}, \sigma_{\textrm{s}, 1}) +
            \alpha_{\textrm{s}, 2} \, \mathcal{N}(\phi_2 \given \mu_{\textrm{s}}, \sigma_{\textrm{s}, 2}) \\ & \quad +
            \alpha_{\textrm{f}} \,
                \mathcal{N}(\phi_2 \given \mu_{\textrm{f}}, \sigma_{\textrm{f}})
    \end{split}
\end{equation}
where $\alpha_{\textrm{bg}}$ is the fraction of background stars in the field, $\alpha_{\textrm{s}} = \alpha_{\textrm{s}, 1} + \alpha_{\textrm{s}, 2}$ is the fraction of stars associated with the mean stream track, and $\alpha_{\textrm{f}} = (1 - \alpha_{\textrm{bg}} - \alpha_{\textrm{s}})$ is the fraction of stars associated with either the spur, the blob, or at the location of the spur/blob in the control field.
% and ignore uncertainties on the sky position $\phi_2$.
}
\changes{
We generate posterior samples in the parameters of this model in each $\phi_1$ range by specifying uniform priors over all parameters except the variances, for which we use priors that are uniform in the logarithm of the variance.
We use an ensemble Markov Chain Monte Carlo (MCMC) sampler (\citealt{Foreman-Mackey:2013}) to generate posterior samples and run the sampler with 64 walkers for a total of 4096 steps, discarding the first 2048 samples as ``burn-in.''
\figurename~\ref{fig:density-model} summarizes the results of this inference: top row shows histograms of the $\phi_2$ positions of stars in each $\phi_1$ range (left to right) with maximum a posteriori inferred density over-plotted (black curve) and bottom row shows marginal posterior probability distributions for $\alpha_{\textrm{bg}}$, $\alpha_{\textrm{s}}$, and $\alpha_{\textrm{f}}$ estimated from the posterior samples for each $\phi_1$ range.
The 15th and 85th percentile values of the feature amplitude in different fields are $(0.12, 0.19)$ (spur), $(0.07, 0.22)$ (blob), and $(0.005, 0.103)$ (control).
The posterior samples for the amplitude of each feature mixture component are consistent with zero in the control field, marginally significant in the blob field, and significantly inconsistent with the background in the spur field.
% In each of the feature fields, the posterior samples for the feature mixture component amplitude are inconsistent with 0, implying that the features are not consistent with the background stellar density.
% In each of the $\phi_1$ regions, the 15th and 85th percentile values of the feature amplitude, $\alpha_{\textrm{f}}$, are $(0.12, 0.19)$ (spur), $(0.07, 0.22)$ (blob), and $(0.005, 0.103)$ (control).
}

Stars from the spur and the blob are similar to those along the main stream track in the color-magnitude diagram and in proper motions, which confirms their likely association to GD-1.
\changes{Both the spur and the blob are close to large gaps in the stream, so a mechanism that displaces stars from their original orbits along the stream to orbits beyond the stream would explain both sets of features.
Future observations of the full 3-D kinematics combined with detailed dynamical modeling of the GD-1 system will be able to test this scenario.}

% The proximity of these off-stream features to large gaps in the stream is suggestive of the gaps being their origin, further implied by hints of relative velocity differences.
% For example, the median proper motion of the blob is offset from the adjacent stream field by $\approx0.3$\,mas\,yr$^{-1}$ in the positive $\mu_{\phi_1}$ direction.
% We defer detailed modeling of these features to infer their co-evolution with the stream for future work.

\section{Discussion}
\label{sec:discussion}

In this work, we map a cold stellar stream in individual stars by combining precise \pans\ photometry with the revolutionary \gaia\ astrometry.
The \gaia\ proper motions were critical for selecting members of the retrograde GD-1 stream, and this success signals a novel way to find members of other structures in the Galactic halo.
In what follows, we discuss how this first view of GD-1 in the \gaia\ era is already transforming what we expect to learn about the Galaxy from streams in general, and GD-1 in particular.

First, the absence of a clear progenitor has been a long-standing hurdle in dynamical modeling of GD-1, mostly due to the unknown true extent of the stream.
Confirming the progenitor location suggested here, $(\phi_1, \phi_2) =
(-13.5^\circ,-0.5^\circ)$, would motivate further searches for stream members
along the trailing arm (positive $\phi_1$) to symmetrically match the extent of
the leading arm.
% for example measuring the (canonical) angle variations along the stream \citep{Bovy:2014}
The stream length is directly proportional to the information it provides on the underlying gravitational potential \citep{Bonaca:2018}, making GD-1 a top modeling priority to map the inner 30\,kpc of the Galaxy, even at the length of $\sim90^\circ$ inferred in this work.
Properly accounting for the progenitor would also allow for studying detailed density structure within the tails, as advocated by \citet{Kupper:2015}.

Next, GD-1 is no longer a simple, one-dimensional structure (see also \citealt{DeBoer:2018}).
Not only are the density variations (gaps) confirmed as highly significant, but for the first time we have mapped tidal debris from a globular cluster beyond the thin stellar track: stars from the GD-1 stream have been found up to $\sim1^\circ$ away from the main stream track.
One explanation for this separated debris could be from velocity substructure within the cluster, but would likely require a fine-tuned orientation to simultaneously explain the sharpness of the neighboring gaps.
% explaining the amount of separation and velocity offset in GD-1 would require the progenitor to rotate at $\gtrsim 10~\kms$, faster than the measured rotation velocity for any known Milky Way cluster (\citealt{}).
% APW note: The more I thought about this, I wasn't sure we could make a very strong claim without running or pointing to some stream simulations with rotating progenitors...
These features are also likely too localized to be explained by chaotic
dispersal from either triaxialilty (\citealt{Price-Whelan:2016}) or the
time-dependent influence of the rotating bar (\citealt{Pearson:2017}).

Encounters between a massive perturber and a thin stream are expected to open gaps in the observed density, and can expel debris into unusual locations and morphologies (e.g., \citealt{Yoon:2011}, see especially \figurename~9).
The encounter origin would be especially exciting: since GD-1 resides in the halo, the only objects massive enough to cross its path are the long-sought-after dark-matter subhalos.

Confirming the origin of gaps and the off-track features in GD-1 will require
both detailed modeling of the stream formation in the presence of massive
perturbers and follow-up radial velocity measurements of candidate GD-1 members.
Radial velocities will help remove the remaining contamination, further
improving the contrast of the stream, and will enable measurements of
differential energy and angular momentum along the stream and off-track
features.
While constraints on subhalo interactions from individual gaps leave serious
degeneracies between, e.g., the time of the encounter and the mass of the
perturber, energy offsets are time-independent and are a promising avenue
towards placing stronger constraints on the nature of the stream and, if
confirmed, the particular pertuber.

% theoretically, we need to understand the co-evolution of the complex stream system under different origin scenarios, while observing radial velocities of candidate GD-1 members would facilitate data interpretation by remove the remaining contamination and enabling studies in the space of conserved quantities.
% -- sentence on getting more from off-track than gaps alone, but somehow more general and not tied specifically to the encounter scenario / or some other nice parting thought

% -- radial velocities, measure 3d velocity structure (unique predictions)
% - if from encounter -- could get more info on the perturber than from the gaps alone, say from delta v -> delta energy
% - cleaner selection (any stars in the gap?)
% - prediction: very negative rv, should be easy to map the full extent

% -- model facets of stream morphology to ascertain if consistent with encounter signatures
% -- so far a lot of focus on gaps produced in encounters -- extra features we detected beside the GD-1 track suggest that further theoretical focus on scattered material is needed to fully understand its dynamical history

% \section{Conclusions}
% \label{sec:conclusions}

% From the combined proper motion and CMD selection of stream members
% (\sectionname~\ref{sec:data}), it is now clear that the GD-1 stream extends
% at least $90^\circ$ in apparent length.
% Several clear under-densities and gaps are also
% \figurename~\ref{fig:sfd-cmd} (middle panel) again shows the proper-motion and CMD selected stream stars, with several features

\acknowledgements{
It is a pleasure to thank
Marla Geha,
David W. Hogg,
Kathryn V. Johnston;
Lauren Anderson,
Vasily Belokurov,
Andrew R. Casey,
Benjamin D. Johnson,
Sergey Koposov,
Mariangela Lisanti,
Edward Schlafly,
and David N. Spergel. % for useful discussions and feedback.

This work has made use of data from the European Space Agency (ESA) mission {\it
Gaia} (\url{https://www.cosmos.esa.int/gaia}), processed by the {\it Gaia} Data
Processing and Analysis Consortium (DPAC,
\url{https://www.cosmos.esa.int/web/gaia/dpac/consortium}). Funding for the DPAC
has been provided by national institutions, in particular the institutions
participating in the {\it Gaia} Multilateral Agreement.  This research was
started at the NYC Gaia DR2 Workshop at the Center for Computational
Astrophysics of the Flatiron Institute in 2018 April.

AB acknowledges generous support from the Institute for Theory and Computation
at Harvard University.
% All code used in this work and all results are available at
% \url{https://github.com/adrn/GD1-DR2}.
}

\software{
    \package{Astropy} \citep{astropy, astropy:2018},
    \package{dustmaps}\footnote{\url{https://github.com/gregreen/dustmaps}},
    \package{gala} \citep{gala},
    \package{IPython} \citep{ipython},
    \package{matplotlib} \citep{mpl},
    \package{numpy} \citep{numpy},
    \package{scipy} \citep{scipy}
}

\bibliographystyle{aasjournal}
\bibliography{gd1}

% \clearpage

% \appendix
% \section{Completeness and the \gaia\ scanning pattern}
% \label{sec:completeness}

% \figurename~\ref{fig:XX} (XX panel) shows the $V$-band extinction
% in the region around the GD-1 stream, computed from the
% Schlegel-Finkbeiner-Davis extinction map (\cite{Schlegel:1998}; hereafter SFD).

% % Notebook:
% \begin{figure}[h]
% \begin{center}
% \includegraphics[width=0.7\textwidth]{nvisits.pdf}
% \end{center}
% \caption{%
% TODO
% \label{fig:TODO}
% }
% \end{figure}

\end{document}